# Magnetic Configurations Related to the Coronal Heating and Solar Wind Generation
# I. Twist and Expansion Profiles of Magnetic Loops Produced by Flux Emergence

Hwanhee LEE
*School of Space Research, Kyung Hee University, 1, Seocheon-dong, Yongin, 446-701, Korea*
*lhhee@khu.ac.kr*
and
Tetsuya MAGARA
*Department of Astronomy and Space Science, School of Space Research, Kyung Hee University, 1, Seocheon-dong, Yongin, 446-701, Republic of Korea*

## Abstract

The generation of outflows from the Sun known as solar winds is coupled with the heating of the solar corona, and both processes are operated in magnetic structures formed on the Sun. To study the magnetic configuration responsible for these processes, we use three-dimensional magnetohydrodynamic simulations to reproduce magnetic structures via flux emergence and investigate their configurations. We focus on two key quantities characterizing a magnetic configuration: the force-free parameter $\alpha$ and the flux expansion rate $f_{ex}$, the former of which represents how much a magnetic field is twisted while the latter represents how sharply a magnetic field expands. We derive distributions of these quantities in an emerging flux region. Our result shows that an emerging flux region consists of outer part where a magnetic loop takes a large flux expansion rate but a small value of $\alpha$ at their photospheric footpoints, and inner part occupied by those loops where a strong electric current flows. We also investigate the expansion profile of a magnetic loop composing an emerging flux region. The profile is given by an exponential expansion type near the solar surface while it is given by a quadratic expansion type in an outer atmosphere. These detailed magnetic configurations obtained by this study contribute to developing a realistic model for the coronal heating and solar wind generation.

## 1. Introduction

In the solar physics there are fundamental problems that remain unresolved. One of them is the presence of the solar corona, a hot atmospheric layer where the temperature reaches a million Kelvin, more than 100 times as high as the temperature at the solar surface. It is now widely believed that the magnetic field plays a crucial role in producing the corona. To maintain such a hot state of the corona against radiative cooling and thermal conduction effects, there must be operated a heating process in the solar atmosphere. To clarify this heating process is the central part of the coronal heating problem (Hollweg 1985; Narain & Ulmschneider 1996; Parker 1994; Aschwanden 2004). There have been proposed two kinds of models to explain the coronal heating: DC (direct current) model and AC (alternating current) model.

In the DC model, an energy injected from the solar surface to the solar atmosphere is once stored inside a magnetic structure as an electric current, and then the dissipation of an electric current directly heats the corona. On the other hand, the AC model assumes a situation where an injected energy is propagated through a magnetic structure in the form of waves, and the dissipation of these waves causes the heating of the corona. In either model, the magnetic configuration of a structure is important because it determines the distribution of an electric current in the structure and the expansion profiles of individual magnetic loops along which those waves are propagated.

The coronal heating is in fact coupled with the generation of continuous outflows from the hot corona (Chapman 1957; Parker 1963). These outflows are known as solar winds and they are classified into two groups, depending on their speeds: fast winds and slow winds. Magnetic configurations producing these outflows have widely been studied, and it is suggested that a fast wind comes from a coronal hole (Wang & Sheeley 1990) while a slow wind may have a relation with an active region formed via the emergence of intense magnetic flux below the solar surface. Baker et al. (2009) show that a quasi-separatrix layer could be a source region of an outflow while Murray et al. (2010) demonstrate that the interaction between an emerging magnetic field and a surrounding vertical magnetic field drives an outflow at the boundary area of an emerging flux region. Recently, it has been reported that a continuous outflow comes from the edge of an active region (Sakao et al. 2007; Harra, et al. 2008; Del Zanna 2008; Doscheck et al. 2008; Marsch et al. 2008). Brooks et al. (2011) suggest that such an outflow may be connected to a slow wind. Furthermore, Riley & Luhmann (2012) point out that a pseudo-streamer, which has a different magnetic configuration from the conventional helmet streamer, may be a source region of a slow wind.

There are two aspects of the coronal heating and the generation of solar winds: i) magnetic configuration and ii) physical mechanism. When we focus on the second aspect, we investigate a physical mechanism for these processes by assuming a simple magnetic configuration. On the other hand, magnetic configurations observed on the Sun are much more complicated than that assumed in the studies of a physical mechanism. It is therefore important to derive the characteristics of the magnetic configuration responsible for the coronal heating and solar wind generation. By deriving them, we could obtain a reliable model for these complex processes.

In the present work we investigate the magnetic configuration of an emerging flux region, which is formed via the emergence of a magnetic flux tube. The emergence of a flux tube injects magnetic energy into the solar atmosphere (Magara 2011), which produces an electric current flowing there. The magnetic configuration of an emerging flux region is characterized by the twist of a pre-emerged flux tube. To see how the twist affects the magnetic configuration of an emerging flux region, we perform separate magnetohydrodynamic (MHD) simulations where different degrees of twist are applied to a pre-emerged flux tube. We focus on two key quantities characterizing the magnetic configuration of an emerging flux tube. The first quantity is

$$\alpha = \frac{|(\nabla \times \mathbf{B}) \cdot \mathbf{B}|}{B^2}, \tag{1}$$

which is known as the force-free parameter of a magnetic field. The spatial distribution of $\alpha$ shows how the electric current generated by a twisted magnetic field is distributed in an emerging flux region. The second quantity has the same dimension as $\alpha$, defined by

$$f_{ex} = -\frac{1}{B}\frac{dB}{ds}, \tag{2}$$

where $B$ is the strength of a magnetic field and $s$ is the length of a magnetic field line. The meaning of this second quantity may be clear when we come to the conservation of magnetic flux in a magnetic flux tube:

$$BA = \Phi_0, \tag{3}$$

where $A$ is the cross sectional area of the flux tube and $\Phi_0$ is the magnetic flux (constant along $s$). By differentiating Equation (3) with $s$, we obtain

$$-\frac{1}{B}\frac{dB}{ds} = \frac{1}{A}\frac{dA}{ds}. \tag{4}$$

The right hand side in Equation (4) indicates that $f_{ex}$ is a quantity representing the local expansion rate of a flux tube. In fact, the expansion rate given by Equation (2) is subject to the direction of a magnetic field line, so we use a modified version of Equation (2), given by

$$f_{ex} = sgn\left(\frac{dZ_b}{ds}\right)\left[-\frac{1}{B}\frac{dB}{ds}\right], \tag{5}$$

where $Z_b$ is the height of a field-line element measured from the solar surface. Equation (5) makes the flux expansion rate always positive when a flux tube expands in the vertical direction no matter whether a magnetic field line is directed upward or downward.

The electric current represented by $\alpha$ may contribute to the coronal heating via the dissipation of itself. The flux expansion rate is also suggested to have a close relationship to the heating of the corona and the generation of solar winds. Suzuki (2006) presents a theoretical interpretation on a relation between the flux expansion rate and the speed of a solar wind. According to a recent model presented in Matsumoto & Suzuki (2012), the dissipation of MHD waves propagated along a sharply expanding flux tube produces a hot corona via a shock heating as well as drives a continuous outflow through a turbulent heating.

The organization of this paper is as follows. The next section describes our simulations. We then present a result from the simulations in Section 3. A comparison between dynamically emerging magnetic fields obtained by the simulations and static magnetic fields obtained by an extrapolation method is also presented in this section. In section 4, we discuss the global magnetic configuration of an emerging flux region and the expansion profiles of individual magnetic loops composing that region.

## 2. Model description

We have performed three-dimensional MHD simulations by solving a set of equations given below:

$$\frac{\partial \rho}{\partial t} + \nabla \cdot (\rho \mathbf{v}) = 0, \tag{6}$$

$$\rho \left[ \frac{\partial \mathbf{v}}{\partial t} + (\mathbf{v} \cdot \nabla) \mathbf{v} \right] = -\nabla P + \frac{1}{4\pi} (\nabla \times \mathbf{B}) \times \mathbf{B} - \rho g_0 \hat{\mathbf{z}}, \tag{7}$$

$$\frac{\partial \mathbf{B}}{\partial t} = \nabla \times (\mathbf{v} \times \mathbf{B}), \tag{8}$$

$$\frac{\partial P}{\partial t} + \nabla \cdot (P \mathbf{v}) = -(\gamma - 1) P \nabla \cdot \mathbf{v}, \tag{9}$$

and

$$P = \frac{\rho \Re T}{\mu}, \tag{10}$$

where $\rho$, $\mathbf{v}$, $\mathbf{B}$, $P$, $g_0$, $\gamma$, $\mu$, $\Re$, and $T$ mean the gas density, fluid velocity, magnetic field, gas pressure, gravitational acceleration, adiabatic index ($\gamma = 5/3$ is assumed), mean molecular weight ($\mu = 0.6$ is assumed), gas constant and temperature, respectively. The units of these physical quantities are listed in Table 1.

The simulation domain is $(-200, -200, -10) < (x, y, z) < (200, 200, 190)$ where the $x$ and $y$-axes define a horizontal plane at the solar surface while the $z$-axis is directed upward. The domain is discretized into grids whose size is $(\Delta x, \Delta y, \Delta z) = (0.1, 0.2, 0.1)$ for $(-8, -12, -10) < (x, y, z) < (8, 12, 15)$ and it gradually increases toward 4 as $|x|$, $|y|$ and $z$ increase. The total number of grids is given by $N_x \times N_y \times N_z = 371 \times 303 \times 353$.

The initial state of the simulations is given by a magnetized plasma in mechanical equilibrium. The plasma is stratified under an uniform gravity, forming a background atmosphere that extends from a subsurface region (Magara 2012). A cylindrical flux tube is placed horizontally below the surface in such a way as the axis of the flux tube is along a line $(x, y, z) = (0, y, -4)$. The magnetic field composing the flux tube is defined by the so-called Gold-Hoyle profile:

$$\mathbf{B} = B_0 \frac{-b\, r\, \hat{\theta} + \hat{y}}{1 + b^2\, r^2}, \tag{11}$$

where $\hat{y}$ and $\hat{\theta}$ indicate the axial and azimuthal directions of the flux tube, while $r$, $B_0$ and $b$ are a radial distance from the axis of the flux tube, the strength of the magnetic field at the axis and the twist of the magnetic field surrounding the axis. The gas pressure inside the flux tube is reduced so that the total pressure equilibrium is maintained at the boundary of the flux tube (Magara & Longcope 2003). In the present simulations the radius of the flux tube is fixed to 2 while we set either $b = 1$ or $b = 0.2$ which corresponds to a strongly twisted (ST) and weakly twisted (WT) case.

The simulations are initiated by applying the following velocity perturbation to the flux tube during $0 < t < t_r$:

$$v_z = \begin{cases} \frac{v_0}{2}\cos\left(2\pi\frac{y}{\lambda}\right)\sin\left(\frac{\pi}{2}\frac{t}{t_r}\right) & \text{for } |y| \le \frac{\lambda}{2} \\ \frac{v_0}{2}\cos\left(2\pi\frac{y-\left[2L-\frac{\lambda}{2}\right]\frac{|y|}{y}}{4L-2\lambda}\right)\sin\left(\frac{\pi}{2}\frac{t}{t_r}\right) & \text{for } |y| \ge \frac{\lambda}{2}. \end{cases} \tag{12}$$

where $\lambda = 30$, $L = 200$, $d = 4$, $v_0 = 0.31$ and $t_r = 5$. This makes a single $\Omega$-shaped emerging part of the flux tube.

We impose a periodic boundary condition at $y = \pm 200$ and free permeable conditions at other boundaries except for the bottom boundary ($z = -10$) where all the physical quantities are fixed to their initial values. To reduce the effect of waves reflected at a boundary on the simulations, we place a wave damping zone near all the boundaries.

Since the goal of the present study is to investigate the evolution and structure of a magnetic field that emerges and continuously expands outward, we adopt a larger simulation domain than the one we used in our previous work (Magara 2012). We reduce the gas pressure in an outer atmosphere, which is favorable for the continuous expansion of an emerging magnetic field (Tajima & Shibata 1997). To do so, we broadened the range in the lower atmosphere along which the temperature remains constant and the gas pressure drops rapidly, so that a smaller coronal gas pressure than in the previous work is obtained (the location where $T = 50T_{ph}$ is shifted from $z = 7.5$ for the previous work to $z = 10$ for the present work).

## 3. Result

### *3.1. Overview of evolution*

Before going into the details of our result, we briefly explain the evolution of an emerging flux tube reproduced by the simulations. Figure 1 shows snapshots of an emerging flux tube in the ST case (left panels) and WT case (right panels). Magnetic field lines are drawn in color while the grey-scale maps represent the magnetic flux at the solar surface (photosphere). The top and bottom panels show the initial state and a late state of the evolution in each case. At the late state, the emergence of a magnetic field almost saturated. The evolution of an emerging flux tube presented here is similar to what has been reported in many works, and a review of these works is found in Shibata & Magara (2011). When a strongly twisted flux tube emerges into the solar atmosphere, a helical structure of a coronal magnetic field (flux rope) is formed (bottom-left panel) while a quadrupolar-like distribution is observed in the photosphere (Magara et al. 2011). On the other hand, in the WT case an emerging magnetic field shows a diverging configuration without flux ropes in the corona (Magara 2006) while a fragmented distribution is observed in the photosphere (bottom-right panel; see also figure 5b).

### 3.2. *Distributions of $\alpha$ (electric current) and flux expansion rate*

In this subsection we show the distributions of $\alpha$ (electric current) and the flux expansion rate in a magnetic structure formed by either the strongly or weakly twisted flux tube.

#### 3.2.1. *ST case*

Firstly we focus on the ST case. Figure 2 presents two-dimensional maps of the flux expansion rate (left panels, linear scale) and the current density strength (right panels, logarithmic scale) at selected atmospheric layers ($z = 0, 2, 6, 10$). Time is $t = 40$. The solid and dashed contours represent positive and negative magnetic flux. While a quadrupolar-like distribution of magnetic flux is observed in the photosphere ($z = 0$), the magnetic field assumes a bipolar distribution at the high atmospheric layers ($z = 2, 6, 10$). A high current density tends to be distributed around a polarity inversion line at each layer, while a large flux expansion rate is distributed in a region where intense magnetic flux exists. In the boundary area where the magnetic flux becomes weak, the current density strength dramatically decreases while the flux expansion rate does not decrease so much.

Top and bottom panels in Figure 3a present three-dimensional distributions of $\alpha$ and the flux expansion rate along emerging field lines. Time is $t = 40$. A perspective view is given in the left panels while a top view is given in the right panels. It is found that inner coronal loops take a relatively large value of $\alpha$, illuminating a double-J shaped structure known as a sigmoid, indicating that a strong electric current flows in the corona. On the other hand, outer coronal loops overlying these inner loops take a small value of $\alpha$ but large flux expansion rate especially around their footpoints. Figure 3b shows a scatter plot indicating a relationship between $\alpha$ and flux expansion rate at the footpoints of the outer loops whose maximum height is larger than 25. This plot suggests that outer loops tend to have a large flux expansion rate but small value of $\alpha$ at their footpoints.

In Figure 3c we selected a typical outer and inner loop to show how the cross sectional area of a loop ($A$) and the current density strength ($J$) vary along these loops. The left panel shows a top view of the outer (violet) and inner (light-blue) loops where colors represent $\alpha$ in the same scale as in Figure 3a. The top-right and bottom-right panels show graphs of $A(s)$ and $J(s)$ where the dotted and dashed line represent the outer and inner loop. Here $s$ is the length measured from a photospheric footpoint marked an asterisk in the left panel. $A$ and $J$ are normalized by their photospheric values at the footpoint ($A_0$, $J_0$). The cross sectional area increases with $s$ up to either more than 100 times for the outer loop and 10 times for the inner loop compared to $A_0$, and then it decreases toward a photospheric value at another footpoint. On the other hand, the current density strength decreases with $s$ in both loops. A relatively high current density (about 1 % of $J_0$) is observed at coronal part of the inner loop while the current density decreases much more in the case of the outer loop.

### *3.2.2. WT case*

Next we investigate the WT case. Figure 4 is similar to Figure 2, presenting two-dimensional maps of the flux expansion rate (left panels) and current density strength (right panels) at selected atmospheric layers ($z = 0, 2, 6, 10$). Time is $t = 61$. The magnetic field at the low atmospheric layers ($z = 0$ and 2) is fragmented compared to the ST case, although the magnetic field shows a bipolar distribution at the high atmospheric layers ($z = 6$ and 10) where a polarity inversion line shows an S shape (or double-S shape) which is opposite to an inverse-S shaped inversion line observed in the ST case. The current density strength has a rather sporadic distribution at the low atmospheric layers while a high current density tends to concentrate around the polarity inversion line at the high atmospheric layers. The flux expansion rate takes a large value in a region where intense magnetic flux exists, decreasing gradually toward the boundary area where the magnetic flux becomes weak. On the other hand, the current density strength dramatically decreases in the boundary area, as is similar to the ST case.

Three-dimensional distributions of $\alpha$ and the flux expansion rate along emerging field lines are presented in the top and bottom panels in Figure 5a. Time is 61. The structure of an emerging magnetic field is divided into two parts; inner part is occupied by short and low loops where a strong electric current flows, which are overlaid by long loops where a large flux expansion rate is found especially around their footpoints. Compared to the ST case, a strong electric current only flows near the solar surface, while a weak current flows in the corona. A scatter plot in Figure 5b shows a relationship between $\alpha$ and the flux expansion rate at the footpoints of outer loops, suggesting a negative correlation between them although the correlation is weaker than in the ST case.

Figure 5c is similar to Figure 3c, showing how the cross sectional area of a loop ($A$) and the current density strength ($J$) vary along selected inner and outer loops. These loops are displayed in the left panel (the outer loop is mostly drawn in violet and the inner loop is in red) while graphs of $A(s)$ and $J(s)$ are given in the top-right and bottom-right panels, respectively. The outer loop expands more than 100 times from the photosphere to the corona, while the inner loop expands just barely. Also, a high current density is distributed all through the inner loop without any significant decrease.

## *3.3. Comparison between dynamically emerging fields and extrapolated static fields*

We then compare the configuration of a dynamically emerging field to that of a static field obtained from an extrapolation method applied to a photospheric magnetic field reproduced by the simulations. The extrapolation of a coronal magnetic field is frequently used in observational studies because until now it is not possible to fully capture a coronal magnetic field only by observations. For the present study, we use a potential-field extrapolation method explained in Magara & Longcope (2003). A potential field is obtained from $t = 40$ data in the ST case and $t = 61$ data in the WT case. Since in a potential field there is no electric current except at the photospheric boundary, we only investigate the distribution of the flux expansion rate in those ST and WT potential fields.

Figures 6 presents two-dimensional maps of the flux expansion rate in the ST (left panels)

and WT (right panels) potential fields. These potential fields show a similar distribution of the flux expansion rate at the high atmospheric layers ($z = 6$ and 10) while the distributions are quite different at the low atmospheric layers ($z = 0$ and 2). Three-dimensional distributions of the flux expansion rate along potential field lines are presented in Figure 7a. The configurations of these potential fields are, in fact, quite different from those of the emerging fields presented in Figures 3a and 5a. Figure 7b shows the variation of the cross sectional area along a selected outer loop in the ST (left panel) and WT (right panel) potential fields. It is found that they tend to have a common expansion profile.

## 4. Discussion

In this section we discuss the characteristics of the magnetic configurations obtained from the simulations. First of all, we would like to mention the so-called *open magnetic field* on the Sun. It is not simple to prove that a field line is completely open using any simulations because their simulation domains are finite. We therefore assume that a magnetic loop could evolve toward an open field when it continuously expands outward. From this viewpoint, outer loops reproduced by the present simulations are regarded as an open field because they tend to expand continuously, as is also demonstrated in Magara & Longcope (2003). On the basis of this argument, we investigate the expansion profiles of several selected outer loops. This may give an important insight into the nature of an open field where a solar wind is generated.

Let us explain two typical types of the expansion profile of a magnetic loop. One is an *exponential expansion* type where the cross sectional area of a loop is given by

$$A(s) \propto e^{as}, \tag{13}$$

where $a$ is constant and characterizes how sharply a loop expands, which is in fact equal to the flux expansion rate:

$$f_{ex}(s) = a. \tag{14}$$

When $a = 0$, a loop does not expand (cross sectional area is constant). The second type is a *quadratic expansion* type where the cross sectional area of a loop is proportional to the square of $s$

$$A(s) \propto s^2, \tag{15}$$

which gives

$$f_{ex}(s) = \frac{2}{s}. \tag{16}$$

We then investigate the expansion profile of an outer loop in the cases of the dynamically emerging fields and the extrapolated potential fields.

Figure 8 shows four graphs where the solid lines represent $f_{ex}(s)$ of four outer loops in a logarithmic scale. The top-left, top-right, bottom-right and bottom-left graph corresponds to an outer loop displayed in the left panel of Figure 3c, left panel of Figure 5c, top-left panel and top-right panel of Figure 7b, respectively. Here $s$ ranges from one footpoint to the top of a loop. The variation of height of a field-line element along these loops $Z_b(s)$ is also given by the dotted lines. The black dashed lines represent the quadratic expansion profile $f_{ex} = 2/s$ given by Eq. (16). The red dash dot lines are fitting curves explained below.

The expansion profiles of the emerging fields are characterized by four different ranges; the first range corresponds to the vicinity of the solar surface ($s$, $Z_b \leq H_{ph}$) where the flux expansion rate is nearly constant, suggesting the exponential expansion of a loop. In the second range ($H_{ph} \leq s$, $Z_b \leq 10H_{ph}$), the flux expansion rate gradually decreases with $s$ and $Z_b$, and then it enters the third range when $s$ and $Z_b$ is over $10H_{ph}$. In this third range, the flux expansion rate shows the quadratic expansion profile. After the third range, the flux expansion rate quickly decreases toward 0, simply indicating that a loop deviates from an open field and assumes a closed-field shape. For an open field, only ranges I - III are expected.

The potential fields are not confined by a surrounding plasma, and so they have a wide range I to expand exponentially. Compared to these potential fields, the range I is short and the range II is prominent in the emerging fields, indicating that these emerging fields are strongly confined by a surrounding plasma so that their expansion is limited near the solar surface (plasma beta is high). The plasma beta then decreases significantly toward the range III where a magnetic field can determine its configuration without being affected by gas pressure. This causes a transition from the range II to range III.

Twist of a magnetic field also contributes to confining a magnetic field, which is suggested by the fact that the decrease of the flux expansion rate in the range II is more prominent in the ST emerging field than in the WT emerging field.

Those key features of the flux expansion rate mentioned above, that is, $f_{ex}(s)$ is almost constant when $s$ is small while it becomes close to $2/s$, may be represented by the following expansion profile:

$$A(s) = A_0\left(1 + \frac{s}{s_0}\right)^2, \tag{17}$$

where $A_0$ and $s_0$ are constant. This gives

$$f_{ex}(s) = \frac{2}{s + s_0}, \tag{18}$$

which indicates that $f_{ex}(s)$ is nearly constant ($2/s_0$) when $s$ is small while it approaches $2/s$ when $s \gg s_0$. In Figure 8, fitting curves based on Eq. (18) are represented by red dash dot lines.

Figure 9 is a summarizing figure of our study. The left panel in Figure 9 schematically illustrates the magnetic configuration of an emerging flux region formed by a strongly twisted (top) and weakly twisted (bottom) flux tube. In the ST case, inner part of an emerging flux region is occupied by twisted loops that form a double-J shaped coronal structure where a strong electric current flows. This structure is observed as a sigmoid, a typical precursor of solar flares (Sterling & Hudson 1997; Matsumoto et al. 1998; Canfield et al. 1999; Titov & Demoulin 1999; Magara & Longcope 2001; Gibson et al. 2002; Pevtsov 2002; Regnier & Amari 2004; Magara 2004, 2006; McKenzie &

Canfield 2008; Archontis et al. 2009; Shibata & Magara 2011). Outer part of an emerging flux region is occupied by continuously expanding loops that have a large flux expansion rate but small value of $\alpha$ at their footpoints. In the WT case, an emerging flux region is also decomposed into inner and outer parts, although the inner part where a strong electric current flows does not form a sigmoidal structure in the corona; rather it forms a group of low loops near the solar surface, which is known as a sea serpent structure (Pariat et al. 2004). The outer part is occupied by continuously expanding loops, which is similar to the ST outer loops, although the WT outer loops rather have a diverging configuration. A typical expansion profile of an outer loop is schematically illustrated in the right panel of Figure 9.

The spatial distributions of $\alpha$ and the flux expansion rate reproduced by the present simulations suggest that magnetic loops are heated via different physical processes in an emerging flux region, that is, outer loops where the flux expansion rate is large but less electric current flows might be heated via the dissipation of waves propagated along a sharply expanding loop, while a direct heating by the dissipation of an electric current could be efficient in inner twisted loops where a strong electric current flows. A large flux expansion rate of a magnetic loop indicates that strong non-uniformity exists in that loop, which may cause the dissipation of propagating Alfvén waves via, for example, a resonant absorption and/or phase mixing process (Aschwanden 2004). This AC heating mechanism may also work in inner twisted loops where the twist of a magnetic field introduces inhomogeneity to the internal structure of a magnetic loop. These arguments, however, are still in a speculative level, so we will further investigate how these two processes, AC and DC heating, are operated in coronal loops with different magnetic configurations.

Regarding the generation of outflows, our result shows that sharply expanding outer loops tend to have their footpoints at the boundary area of an emerging flux region. If, then, an outflow may be generated in such a sharply expanding loop as suggested by Matsumoto & Suzuki (2012), our result seems to be consistent with a recent observational result that an outflow preferentially comes from the edge of an active region (Sakao et al. 2007; Harra, et al. 2008).

Finally, the potential-field extrapolation presented here may reproduce the expansion profile of a loop in a coronal region (range III), although it cannot reproduce a magnetic configuration in the range II where plasma heating and outflow generation could be operated. An advanced extrapolation method such as nonlinear force-free field extrapolation (Inoue et al. 2012) may be useful for reconstructing a magnetic configuration in the range II. An appropriate treatment of the range II is indispensable for a complete understanding of the coronal heating and solar wind generation.

The authors wish to thank the Kyung Hee University for general support of this work. They also appreciate useful comments given by a referee. This work was financially supported by Basic Science Research Program (2013R1A1A2058705, PI: T. Magara) through the National Research Foundation of Korea (NRF) provided by the Ministry of Education, Science and Technology, as well as the BK21 program (20132015) through the NRF.

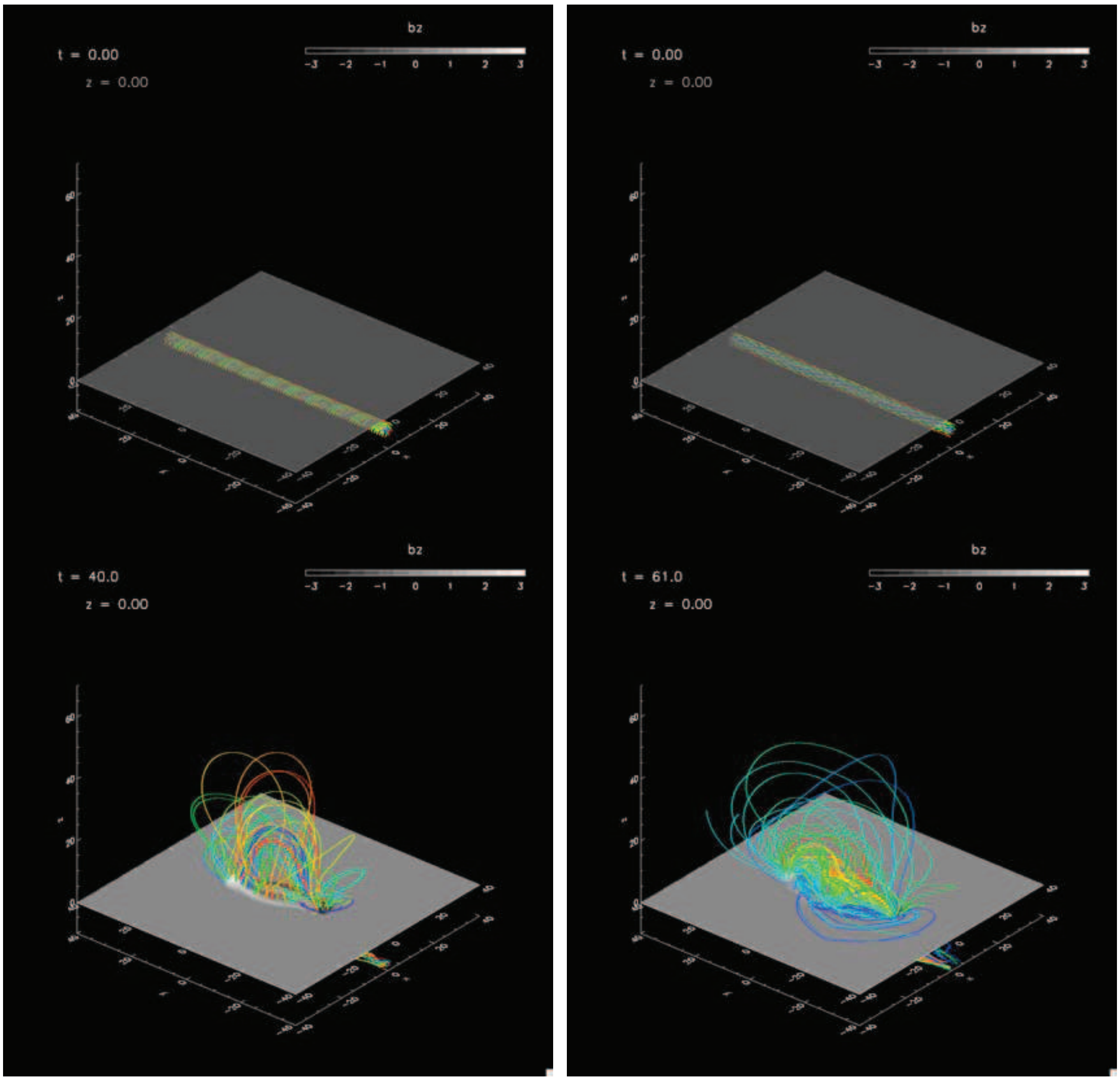

ST case

WT case

**Fig. 1.** Snapshots of an emerging flux tube in the ST case (left panels) and in the WT case (right panels). The color lines indicate magnetic field lines. At $t = 0$ a magnetic flux tube is placed below the solar surface (photosphere) represented by a grey-scale map of magnetic flux. At a late state, a helical structure of emerging magnetic field is formed in the ST case ($t = 40$) while a diverging magnetic configuration is found in the WT case ($t = 61$).

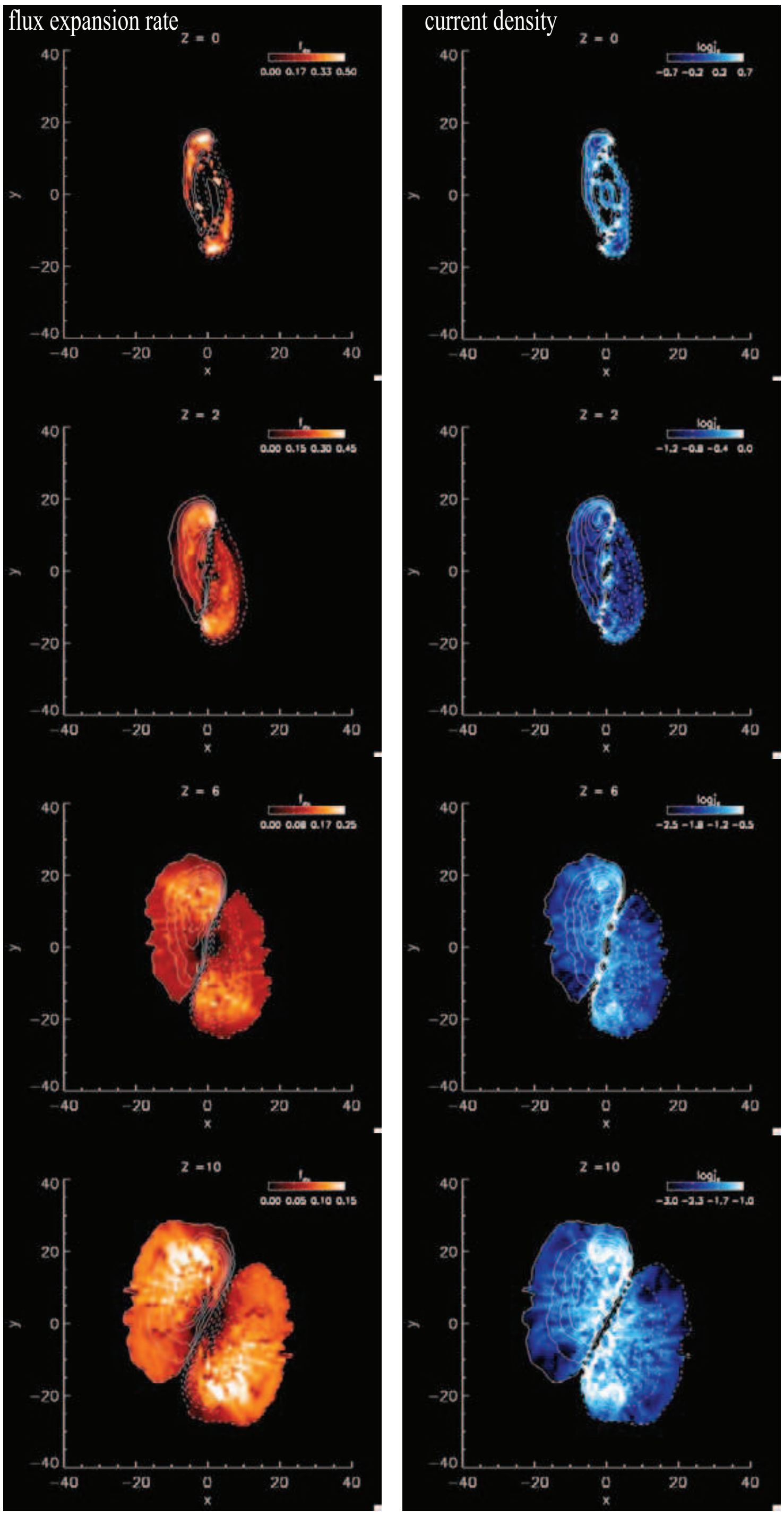

**Fig. 2.** Two-dimensional maps of flux expansion rate (left panels, linear scale) and current density strength (right panels, logarithmic scale) in the ST case are presented at several selected atmospheric layers ($z = 0, 2, 6, 10$). Time is 40. The solid and dashed contours indicate positive and negative magnetic flux, respectively.

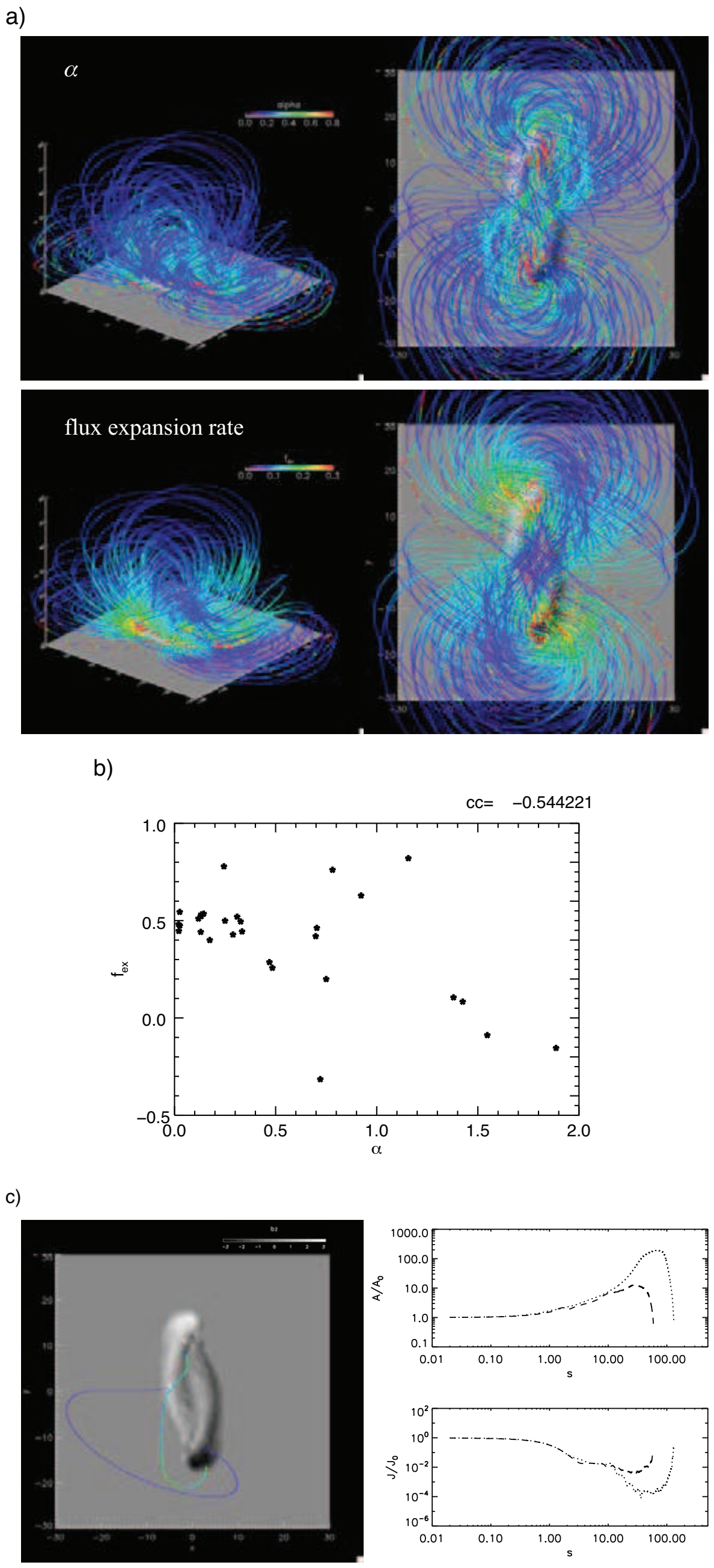

**Fig. 3.** (a) Three-dimensional distributions of force-free parameter $\alpha$ (top panels) and flux expansion rate (bottom panels) along emerging field lines in the ST case are presented. Time is 40. Perspective and top views are given in the left and right panels, respectively. The grey-scale map in each panel represents photospheric magnetic flux. (b) A scatter plot of $\alpha$ and flux expansion rate measured at the footpoints of outer loops (maximum height is over 25) is presented. The correlation coefficient is given at the top-right corner. (c) A top view of selected outer (violet) and inner (light blue) loops is presented in the left panel. Colors represent $\alpha$ in the same scale as in Figure 3a. The grey-scale map shows photospheric magnetic flux. The variation of cross sectional area and current density strength along the outer loop (dotted line) and inner loop (dashed line) is presented in the top-right and bottom-right panel, respectively.

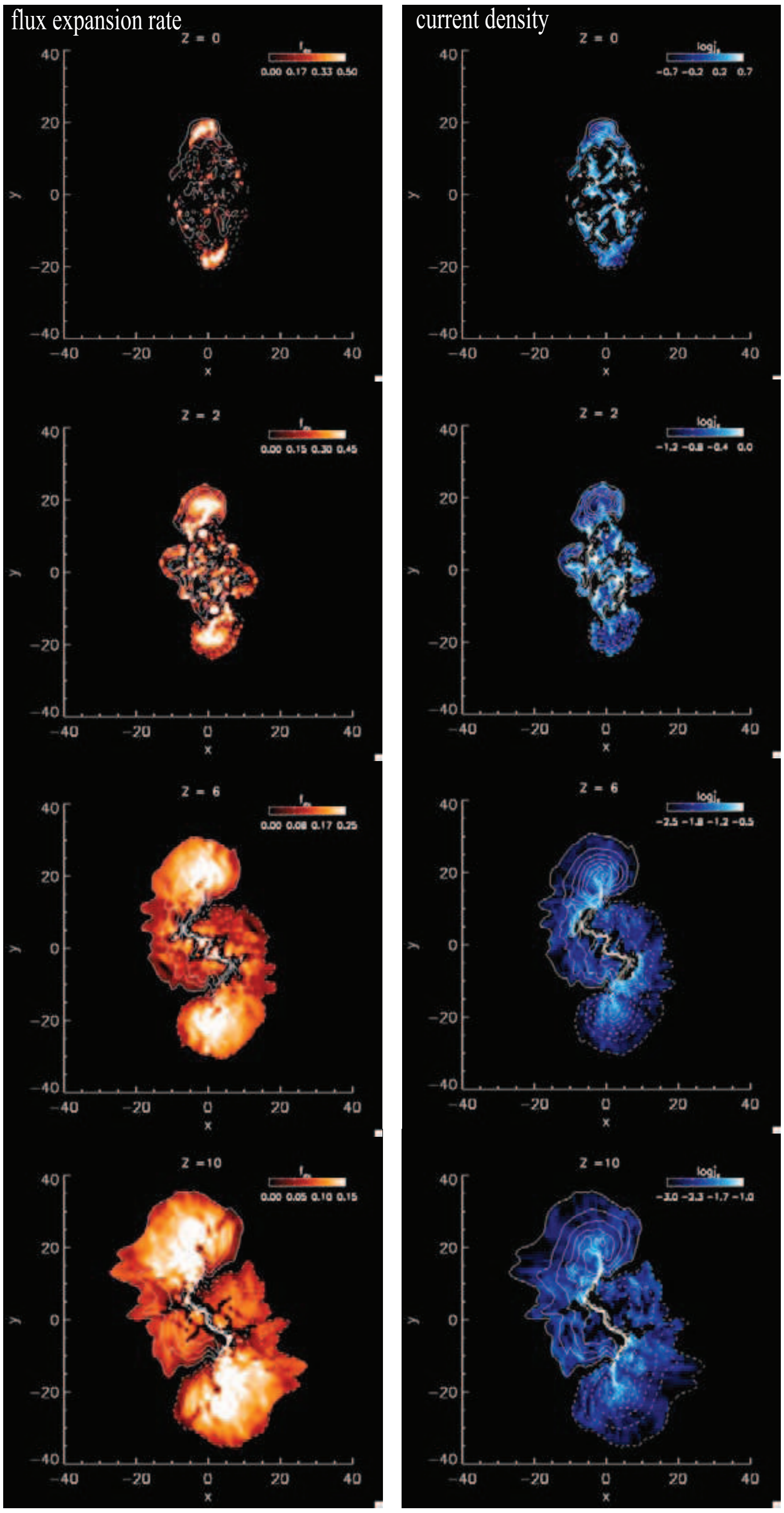

**Fig. 4.** Same as Figure 2 except for the WT case. Time is 61.

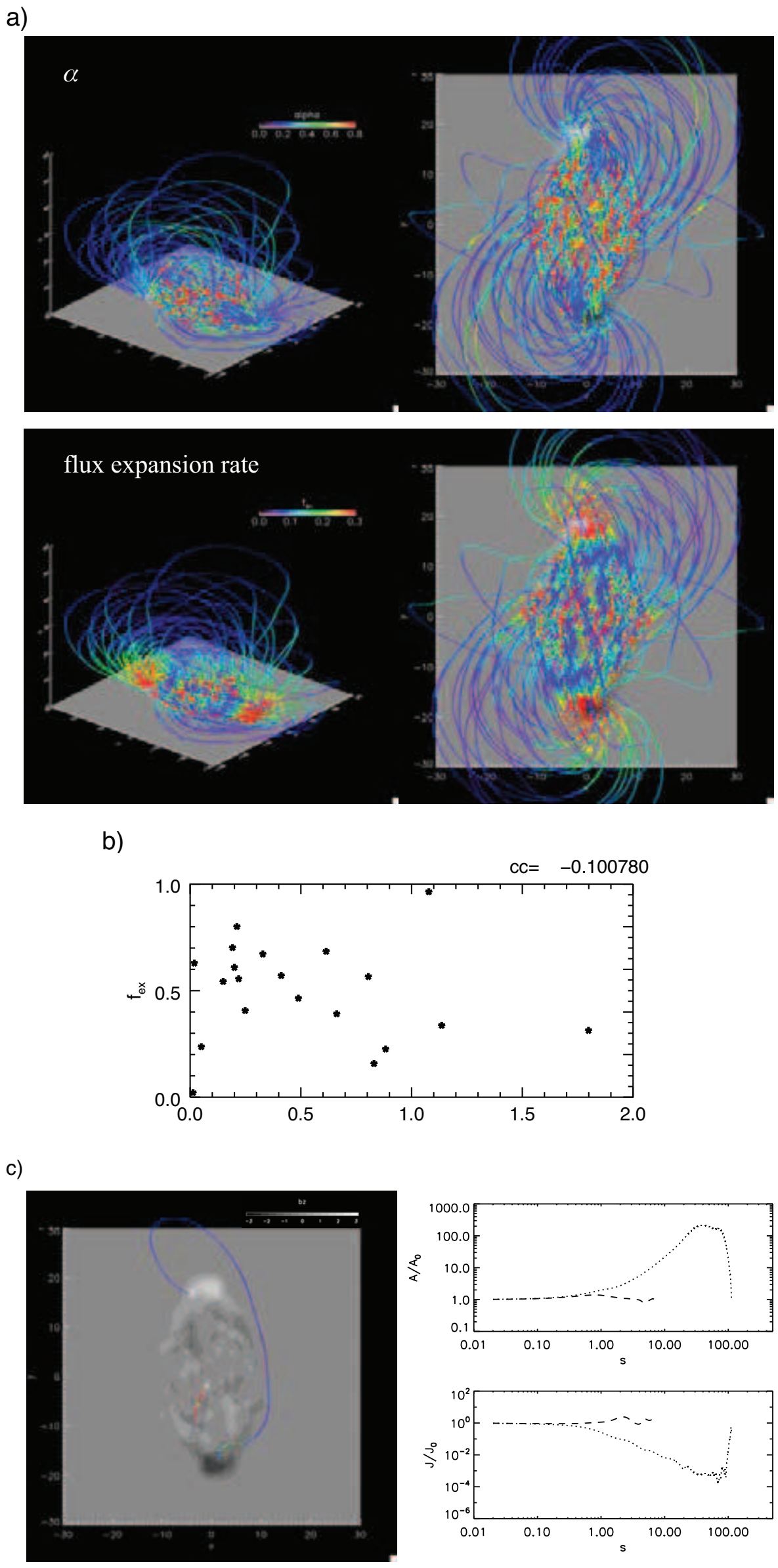

**Fig. 5.** Same as Figures 3abc except for the WT case. Time is 61. In Figure 5b, outer loops whose maximum height is over 10 are plotted.

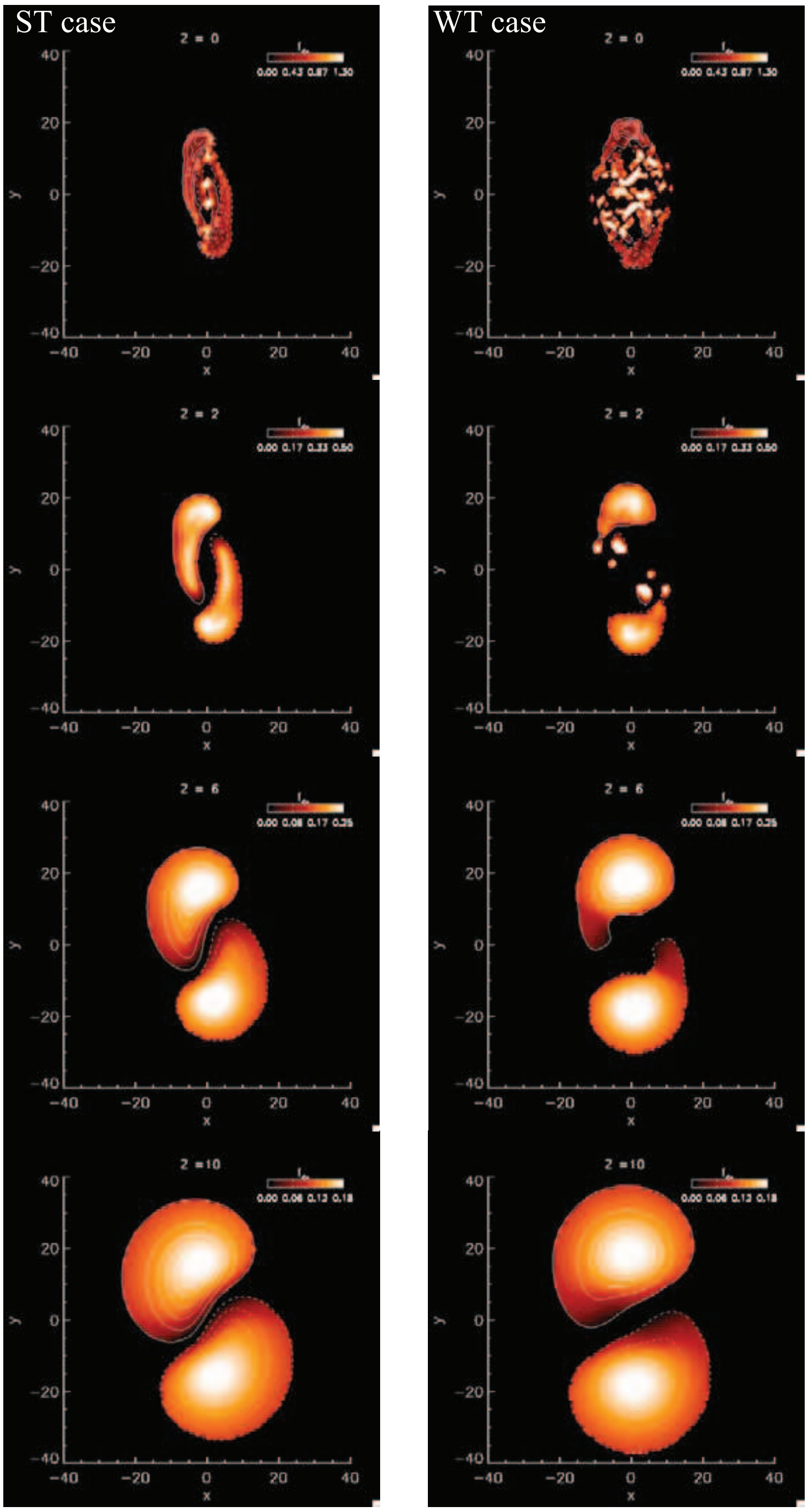

**Fig. 6.** Two-dimensional maps of flux expansion rate in ST potential field (left panel) and WT potential field (right panel) are presented at several atmospheric layers ($z = 0, 2, 6, 10$). The solid and dashed contours indicate positive and negative magnetic flux, respectively.

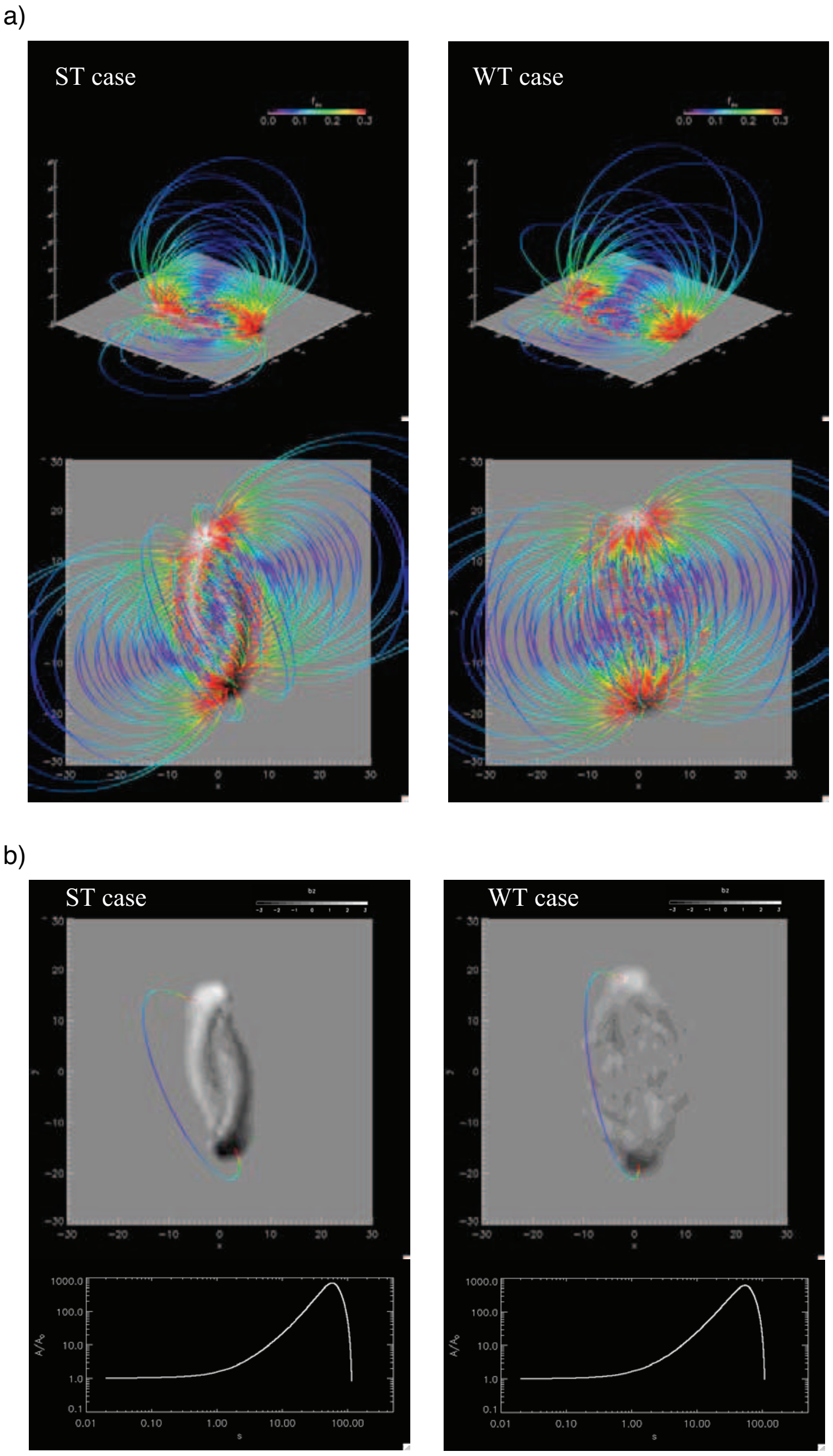

**Fig. 7.** a) Three-dimensional distributions of flux expansion rate along ST potential field lines (left panel) and WT potential field lines (right panel) are presented. Perspective and top views are given at the top and bottom panels. The grey-scale map in each panel represent photospheric magnetic flux. (b) A top view of a selected outer loop is presented in the top-left (ST case) and top-right (WT case) panels. Colors represent flux expansion rate in the same scale as in Figure 7a. The grey-scale map in each panel represents photospheric magnetic flux. The variation of cross sectional area along the outer loop is presented in the bottom-left (ST case) and bottom-right (WT case) panels.

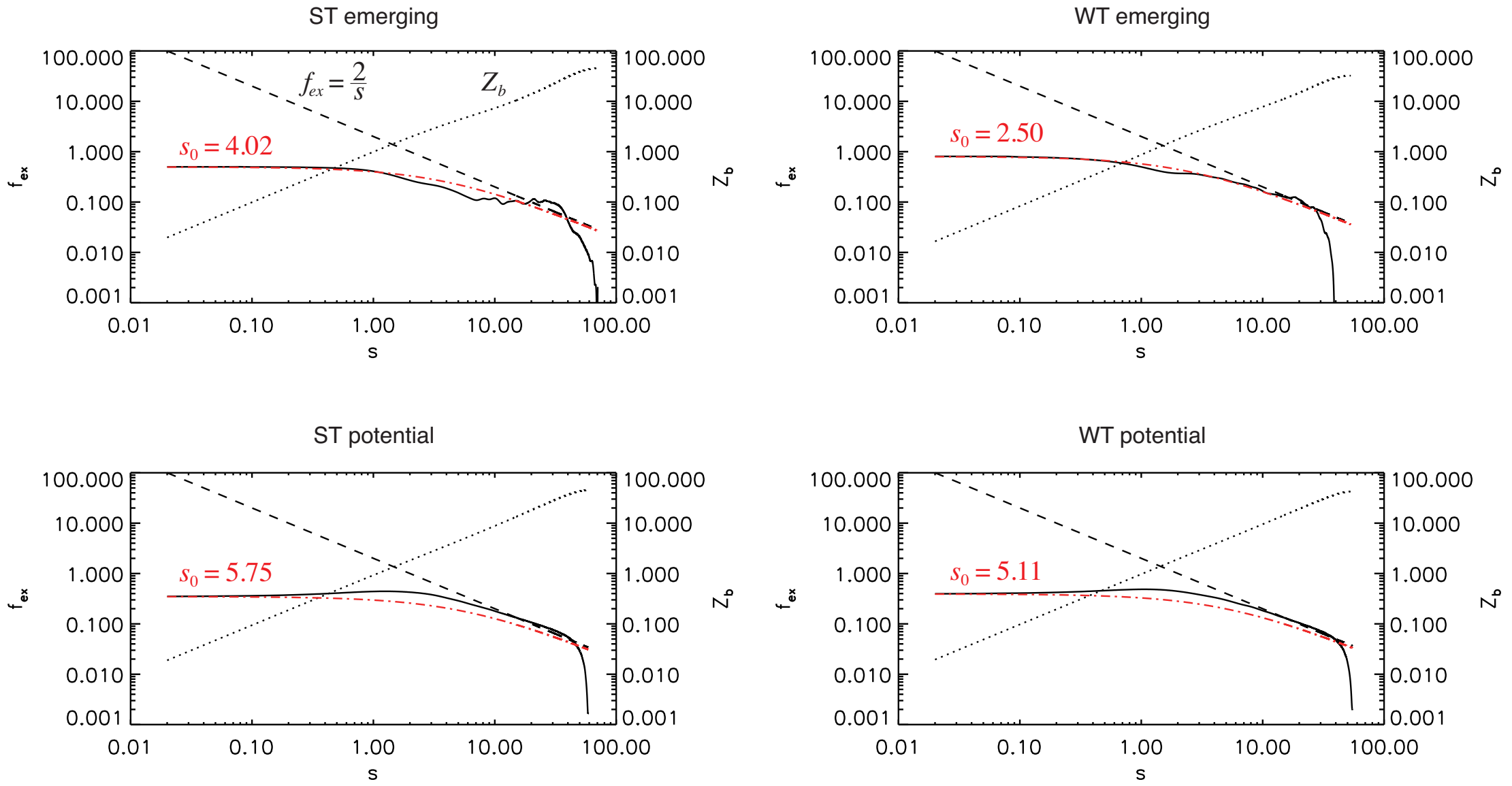

**Fig. 8.** Expansion profiles of four outer loops shown in Figures 3c, 5c and 7b are presented. The solid lines show the variation of flux expansion rate with loop length while the dashed lines indicate the quadratic expansion profile given by Eq. (16). Fitting curves based on Eq. (18) are represented by red dash dot lines. The variation of height ($Z_b$) with loop length is also given by dotted lines. The unit of length is given by $2H_{ph}$ where $H_{ph}$ is photospheric pressure scale height.

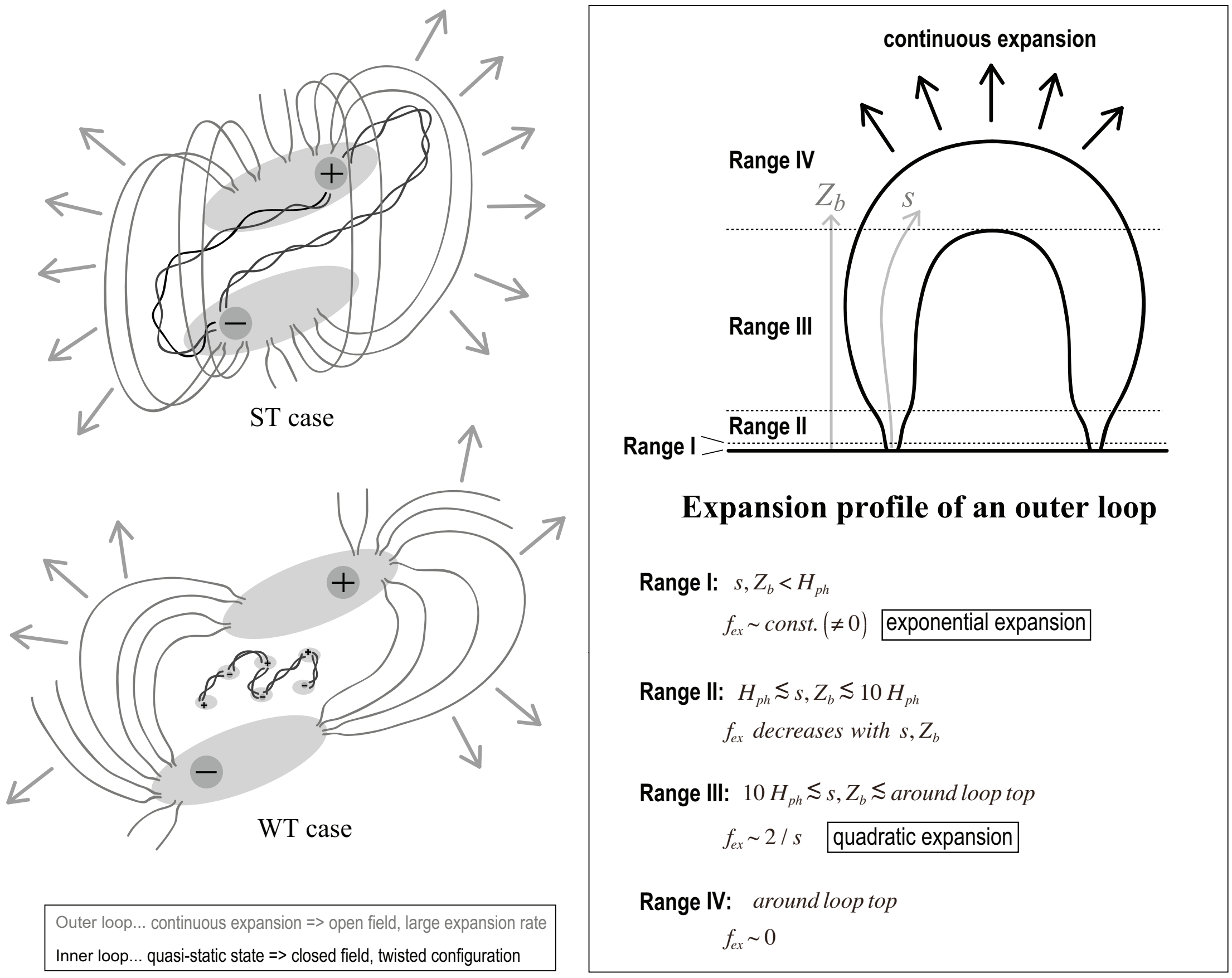

**Fig. 9.** Left panel: Magnetic configuration of an emerging flux region in the ST (top) and WT (bottom) case is schematically illustrated. Right panel: Typical expansion profile of an outer loop is presented. $s$ and $Z_b$ represent the length and height of the loop. $H_{ph}$ is photospheric pressure scale height. For an open field, only ranges I - III are expected. For details, see the text.

**Table 1.** Units of Physical Quantities

| Physical Quantity | $Unit$ |
|---|---|
| Length | $2\ H_{ph}$ [1] |
| Velocity | $c_{s_{ph}}$ [2] |
| Time | $2\ H_{ph}\ /\ c_{s_{ph}}$ |
| Gas Density | $\rho_{ph}$ [3] |
| Gas Pressure | $\rho_{ph} c_{s_{ph}}^2$ |
| Temperature | $T_{ph}$ [4] |
| Magnetic Field | $(\rho_{ph} c_{s_{ph}}^2)^{1/2}$ |

[1] Photospheric gas pressure scale height.

[2] Photospheric adiabatic sound speed.

[3] Photospheric gas density.

[4] Photospheric temperature.